# Conductivity of Dirac-like surface states in correlated honeycomb transition metal oxide Mott insulators

Thomas Dziuba[1], Ina-Marie Pietsch[2], Máté Stark[1], Georg A. Traeger[1], Philipp Gegenwart[2] and Martin Wenderoth[1]

[1]IV. Physical Institute, Georg-August University Göttingen, Friedrich-Hund-Platz 1, 37077 Göttingen, Germany
[2]Experimental Physics VI, Center for Electronic Correlations and Magnetism, Augsburg University, Universitätsstraße 1, 86159 Augsburg, Germany

Date: 26. May 2020

Abstract:

The search for materials with novel and unusual electronic properties is at the heart of condensed matter physics as well as the basis to develop conceptual new technologies. In this context, the correlated honeycomb transition metal oxides attract large attention for both, being a possible experimental realization of the theoretically predicted magnetic Kitaev exchange and the theoretical prospect of topological nontriviality. The Mott insulating sodium iridate is prototypical among these materials with the promising prospect to bridge the field of strongly correlated systems with topology, finally opening a path to a wide band gap material with exotic surface properties. Here, we report a profound study of the electronic properties of ultra-high-vacuum cleaved surfaces combining transport measurements with scanning tunneling techniques, showing that multiple conductive channels with differing nature are simultaneously apparent in this material. Most importantly, a V-shaped density of states and a low sheet resistance, in spite of a large defect concentration, point towards a topologically protected surface conductivity contribution. By incorporating the issue of the addressability of electronic states in the tunneling process, we develop a framework connecting previous experimental results as well as theoretical considerations.

Condensed Matter Physics, Strongly Correlated Materials, Topological Insulators

## I. INTRODUCTION

Correlated honeycomb transition metal oxides (TMO's) $A_2IrO_3$ (A= Li, Na) attract large attention for being at the intersection of various physical phenomena that are in the focus of current research due to their potential relevance for future technological application as well as for fundamental questions. The crystal consists of alternatingly stacked $Na_3$ and $NaIr_2O_6$ layers (see Supplemental Material at [Publisher-URL] for figure S1(a)-(c)), where the structure of the $NaIr_2O_6$ layer with a honeycomb coordination is determined by Iridium atoms surrounded by edge-sharing oxygen octahedra having the basic ingredients of a possible realization of the magnetic Heisenberg-Kitaev exchange interaction [1-6]. The electronic properties are highly complex, as crystal field splitting, strong spin-orbit coupling and Hubbard repulsion have similar energy scales in $Na_2IrO_3$ [1,6-14]. This results in a robust $j=1/2$ Mott-insulating behavior [8] with a bandgap comparable to common semiconductors [8,15,16]. The magnetic properties of $Na_2IrO_3$ are dominated by the strongly bond-dependent Kitaev exchange, although the admixture of Heisenberg and off-diagonal interactions beyond nearest neighbors leads to a zigzag antiferromagnetic ordered ground state [2,17-19]. Furthermore, $Na_2IrO_3$ is theoretically considered as candidate material for correlated topological insulator behavior in two [9-11,20,21] or three dimensions [20]. The observation of topologically non-trivial properties would render $Na_2IrO_3$ to be the first Mott insulating oxide topological insulator [22-29].

To this end, many groups have investigated the surface properties of $Na_2IrO_3$, but no consistent picture has been obtained from the experimentally determined electronic and transport data. Optical conductivity, angle-resolved photoemission spectroscopy (ARPES) and scanning tunneling spectroscopy (STS) investigations reported a wide range of band gaps between 340 meV and 1.2 eV [8,15,16]. Scanning tunneling microscopy (STM) and spectroscopy experiments on freshly cleaved $Na_2IrO_3$ crystals found a bandgap of 1.2 eV and 0.6 eV, associated with a $Na_3[(1x1)]$- and $NaIr_2O_6[(\sqrt{3}x\sqrt{3})R30°]$ surface reconstruction respectively [15]. µ-ARPES investigations confirmed different band gap widths for the two surface terminations [16]. However, the extracted absolute values are not consistent with the STS investigations. Only recently, first indications of in-gap states have been reported from ARPES measurements [30] which are discussed as exotic surface metallicity. Transport investigations have reported a freeze-out of conductivity at low temperature [16,31]. Investigations using conductive AFM under ambient condition showed the formation of sodium clusters at the surface of $Na_2IrO_3$ when exposed to air for several hours, resulting in metallic conductivity on the surface which was attributed to the sodium clusters [32]. In summary, besides the in-gap states observed by ARPES [30], neither a clear band gap closing by surface states in spectroscopic data nor surface conductivity has been found, i.e. typical hallmarks of topologically nontrivial behavior have not been observed so far.

In this work, we combine temperature-dependent transport measurements with scanning tunneling microscopy and spectroscopy investigations at room temperature on freshly cleaved $Na_2IrO_3$ single crystals.

The samples were grown using solid-state synthesis, resulting in thin plate-shaped crystals with a diameter up to 4 mm. The preceding polycrystalline material was grown and characterized as described in former works [14,31]. After adding 10% extra Ir-powder, the mixture was heated up to 900°C, cooled and ground. Afterwards, the product was heated to 1050°C. This results in small single crystals. Larger crystals were obtained by an additional annealing with slow heating to 1050°C followed by a long hold time. Magnetic susceptibility measurements revealed a sharp signature at $T_N$, confirming good crystalline quality.

For macroscopic bulk- and surface conductivity measurements, four gold contacts were attached to the plate-shaped $Na_2IrO_3$ crystals: two at the bottom (between ceramic substrate and sample) and two at the top, using evaporation of high purity gold and shadow masks. The contacts of each pair (top/bottom) were positioned at laterally opposite sides of the crystal. To achieve a clean surface, crystals were cleaved under ultra-high vacuum (UHV) conditions. Temperature depended I-V-measurements were done using a Keithly 2602B source measure unit.

The microscopic measurements were performed in a homebuilt ultra-high vacuum STM at pressures below $5*10^{-10}$ mbar at 300 K. To achieve clean surfaces, the crystals were cleaved in UHV and directly transferred to the STM probing stage. STM topography images were acquired using constant current mode. All voltages refer to the sample bias voltage with respect to the tip, all currents refer to the tunneling current between tip and sample. To gain full scanning tunneling spectroscopy maps dI/dV-spectra were recorded at every second pixel in the topography using lock-in techniques for enhanced resolution. Additionally, I-V-curves were recoded without lock-in techniques and used for the calibration of the dI/dV-curves. The absolute calibration error can be estimated to < 0.5 pA/V at free surfaces and < 2.6 pA/V at step edges.

The used tungsten STM-tips were prepared by electrochemical etching in a KOH-solution, and further processed by annealing and argon sputtering in UHV. During measurement, in-situ modifications by Na occur by chance due to the high sodium mobility on the $Na_2IrO_3$ surface. Intentional cleaning of the tip is done by applying voltage pulses up to 5 V.

Acquired data was analyzed and visualized using the home build software NextDiagram ver. 5.2.1. Visualization of the $Na_2IrO_3$ crystal structure was done using VESTA ver. 3.2.1.

Performing all measurements under ultra-high vacuum conditions, we propose a comprehensive model regarding the electronic properties including the conductive behavior of the $Na_2IrO_3$ surface. Our approach ensures the inclusion of the local surface structures and possible degradation in the discussion. We have observed conductivity on the Na$_2$IrO$_3$ surface in transport measurements as well as an in-gap local density with a linear dispersion of the surface states. Local spectroscopy allows us to disentangle different conductivity channels in dependence of the local surface structure and modifications of the STM-tip, enabling us to classify unusual conductive behavior inside the bulk band gap.

## II. HIGH SURFACE CONDUCTIVITY OF NA$_2$IRO$_3$ AND ITS MICROSCOPIC ADDRESSABILITY

The high reactivity of Na$_2$IrO$_3$ leads to the rapid degradation of its surface [33]. We avoid this problem by performing the temperature-dependent transport measurements in an UHV setup where the single crystals are cleaved in-situ. Before cleaving, the bulk properties of our sample have been acquired (Fig. 1(d)) showing the well-known freeze-out [16,31] of conductivity at lower temperature. Fitting these curves with a simple thermal excitation Ansatz R ~ exp($\Delta E/2k_BT$) for high temperatures (> 240K) yield excitation energies of $\Delta E_{AB}$ = 195meV and $\Delta E_{CD}$ = 205meV for the channels A-B and C-D respectively. It is worth noting, that these values only serve as a lower boundary, since the previously reported hopping mechanism [31] contributes to the overall conductivity of the material which is not reflected by the thermal activation calculation.

After cleavage, we observe a significant increase of the conductivity induced by the surface (Fig. 1(e)), while the conductivity of the bulk remains unaffected. This surface conductivity manifests in a plateau-like resistivity below around 200 K without any indication of a freezing out of conductivity. This resembles the recently reported findings for the electronic transport on the surface of three-dimensional topological insulators [34]. The average resistance of around 10 kΩ on the plateau results in a sheet resistance between 0.5 kΩ/□ and 2 kΩ/□, depending on the estimated contact area between the gold contacts and the sample surface that were evaluated using an optical microscope. In a final step, we expose the surface for 13 hours to air. This results in a surprisingly slow degradation of the surface manifesting as a reduction of the surface conductivity (Fig. 1(f)). In contrast to previous resistivity measurements, which displayed insulating behavior at low temperature [16,31], our data show a clear indication of a surface-related conducting channel for a freshly cleaved sample.

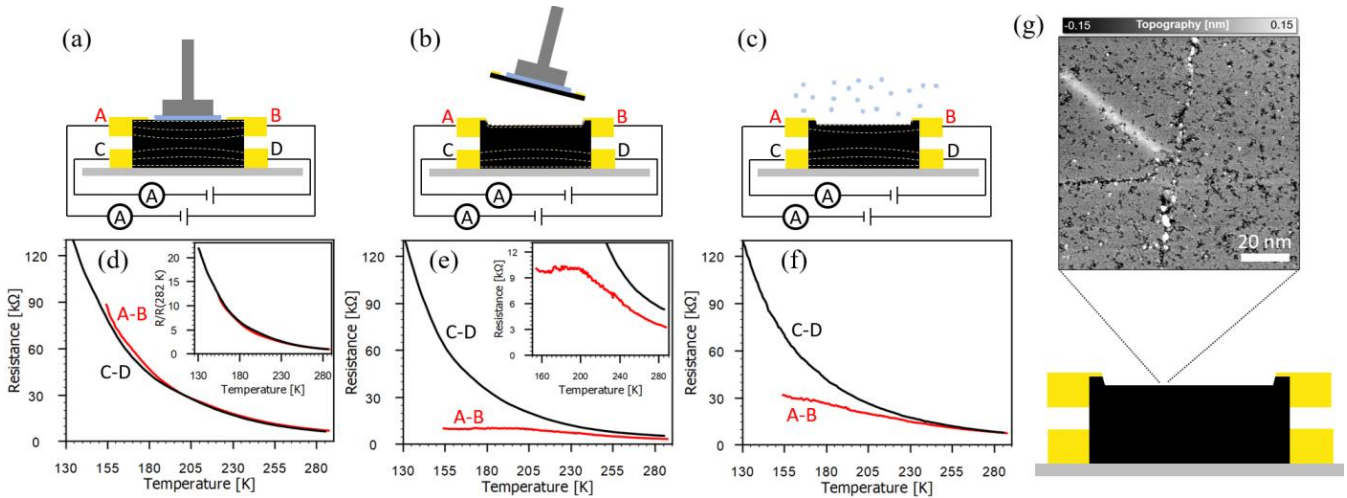

FIG. 1. Surface conductivity of cleaved Na$_2$IrO$_3$. To distinguish surface-related conductivity from bulk conductivity, four contacts were attached to the sample: two for contacting the (freshly cleaved) crystal surface (A-B) and two for reference measurements through the uncleaved crystal (C-D). (a) Before cleavage, measurements of the bulk conductivity were done for both channels. The yellow dashed lines mark the anticipated current flow. (d) The conductivity of the bulk crystal shows a strong temperature dependence between 285K and 155K. The experimental results can be modeled with thermal activation at higher temperatures, resulting in excitation energies of 195 meV and 205 meV for the channels A-B and C-D, respectively. (b) After cleaving in UHV, (e) the conductivity at low temperatures is completely dominated by the surface, reducing the resistivity by nearly an order of magnitude at 155K. (c) After exposing the surface by air for several hours, (f) the surface conductivity is reduced but not completely suppressed. (g) 100nm x 100nm constant current

topography of the freshly cleaved surface. Defects ranging from vacancies to cracks and ridge-like defects are always apparent, making the high conductivity (see (d)) even more remarkable. For further topographic details see Supplemental Material [Publisher-URL] fig. S2.

### III. SPECTROSCOPIC CHARACTERIZATION OF $NA_2IRO_3$ SURFACE STATES

Further structural and electronic investigations have been done using STM and STS. In constant current topography measurements on the freshly cleaved surface we observe the known (1x1)- and ($\sqrt{3}$x$\sqrt{3}$)R30°-terminations. In contrast to investigations under ambient conditions [32], we neither observed degeneration of the surface nor a modification due to a diffusion of sodium forming clusters. However, as can be seen from figure 1(d), one can find many different defects on the surface. As large atomically ordered areas are mainly found for the (1x1)-reconstruction, we have concentrated our investigation on this reconstruction.

Figure 2 shows that a significant in-gap local density of states is also observed in STS. In addition, such data show a strong spatial variation of the differential conductivity. The simultaneously acquired constant current topography of the (1x1)-reconstructed surface, depicted at the bottom of figure 2, displays a drop in the topographic height by 0.43 nm at the position marked with a black arrow, reflecting a tip modification. Our measurements have shown that such attachments and detachments of sodium to the probe are common in STM studies on $Na_2IrO_3$. Since the slow scan axis is oriented from top to bottom, we propose that a sodium atom attached to the tip was disposed in this process. Comparing the topographic and spectroscopic data, it becomes clear that the tip modification results in a sudden change in the dI/dV-maps, emphasizing the role of tip modifications in regard to the addressability of electronic sample states. For the following discussion, it is important to note that we keep atomic resolution across the tip modification (see Supplemental Material [Publisher-URL] fig. S3).

To further analyze and discuss the spectroscopic in-gap information, we have classified our dI/dV-curves according to three generic types of spectra. Examples of these spectra are depicted in figure 3(a)-(c).

Type 1 spectra show a clear band gap of 0.8 eV close to previous STM investigations [15]. We refer to this gap between valence band edge at -0.45 eV and the conduction band edge at 0.35 eV in the following as the "bulk band gap", being aware that this value was found to be different on the two surface reconstructions [15]. Most importantly, these type of spectra shows no hallmark of in-gap local density of states.

Type 2 spectra exhibit additional peaks at -0.5 eV and 0.5 eV close to the valence and conduction band. These peaks are symmetrically broadened over several hundred meV each. Their in-gap conductivity has a minimum at the Fermi level and a non-linear dispersion. The onset of tunneling into the valence band can be identified.

Type 3 spectra paint a very different picture: the band gap is entirely closed in a V-shaped fashion, with the minimum at the fermi level. The onset of tunneling into the conduction band states is visible as a tiny shoulder.

At this point, it is important to note that from data sets like shown in figure 2 we conclude that the spectral shape of the in-gap density of states is highly dependent on both, the local electronic surface properties and the electronic properties of the STM-tip. Datasets showing typical examples of different types of conductivity are depicted in figure 3(d)-(k), as well as in the Supplementary Material [Publisher-URL] figures S4 and S5.

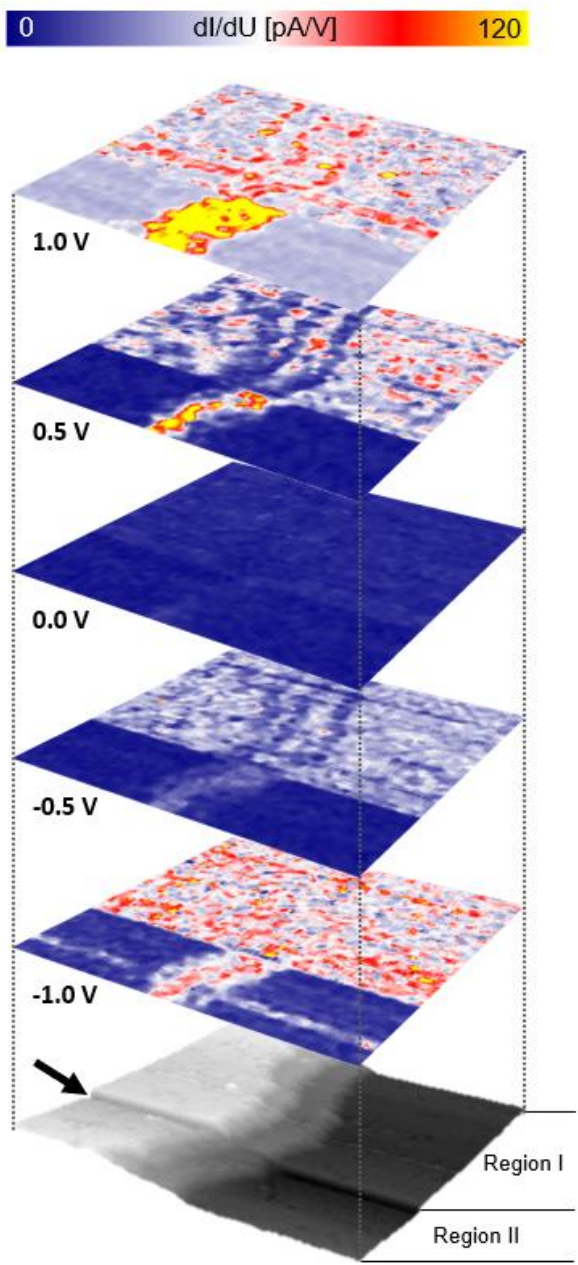

FIG. 2. Spatially resolved in-gap conducting channels on a stepped $Na_2IrO_3$ surface. dI/dV-curves have been acquired at every second pixel of a constant current topography (shown in greyscale). The measurement covers an area of 40x40 nm with three single steps. All surfaces are (1x1)-reconstructed (see Supplementary Material figure S3). An in-situ tip modification can be recognized by an immediate jump in the topographic height, marked by a black arrow. Conductivity dI/dV-maps labeled according to the applied bias voltage cover the energy range of $Na_2IrO_3$ from the valence band edge across the band gap to the conduction band. A clear in-gap current is found for bias voltages between -0.5 V and 0.5 V. Both, the step edges and the modification of the tip are clearly visible in the dI/dV-maps. Region I and region II label the areas examined before after the tip modification, respectively. In region I, lateral fluctuations of the dI/dV-value on free surfaces are supplemented by conductivity alterations at the step edges. In region II, the overall conductivity on the highest and the lowest terrace is reduced by an order of magnitude, where the terrace in the center shows enhanced conductivity. Here, step edges do not provide a characteristic alteration of the conductivity in contrast to region I.

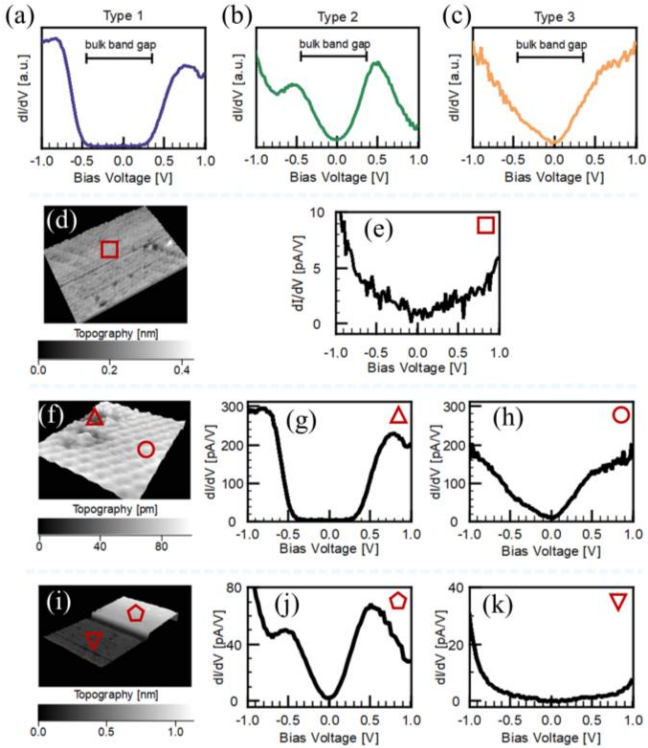

FIG. 3. Three generic types of spectra are observable. All acquired spectra can be sorted according to simple criteria. (a) Type1: The bulk band gap of 0.8 eV is very pronounced with no residual in-gap conductivity. This spectrum represents the expected contribution from bulk conductivity. (b) Type 2: In-gap differential conductivity showing broad, Gauss-shaped peaks at -0.5 V and 0.5 V bias voltage, resembling e.g. conductivity via defect bands. (c) Type 3: These dI/dV-spectra are characterized by a V-shaped band gap closing and a minimum of conductance close to the Fermi level. The onset of tunneling out of the valence and into the conduction bulk band is visible as small shoulders. The contribution of the different channels to the tunneling current strongly depends on the local electronic structure of the surface, but also on the tip properties. (d)-(e) Type 3 spectra are observed on the (1x1)-surface. (f)-(h) While type 1 spectra are found on the (1x1) surface (g), a type 3 spectrum is observed at the defect (h). (i)-(k) After a tip modification, while maintaining atomic resolution, a shift in the spectral weight from type 2 (j) to a superposition of type 1 and type 3 (k) is observable.

## IV: PHYSICAL CLASSIFICATION OF THE SPECTROSCOPIC RESULTS

Following the idea of Singh *et al.* [31], the bulk conductivity can be explained by thermally activated charge carriers at high temperatures (200 K - 300 K) and hopping transport at lower temperatures. Modeling our findings accordingly results in activation energies in the range of 200 meV. Our STS data suggest that the Fermi Energy is - at least at the surface - located close to the center of the band gap. Since the bulk band gap taken from STS is much larger (0.8eV, see figure 3(a),(g)), we propose that the activation energy can be attributed to the difference between defect states and the valence and the conduction band edge. The presence of defects is directly shown by the topographic information from STM. Moreover, for the observed high defect concentration, in-gap conductivity peaks as observed as type 2 can be associated with defect bands, an established concept in semiconductors. Taking the energetic center of these defect-related bands, the difference to the band edges is close to the activation energy determined from the bulk transport data. Presuming the density of defects found at the surface continues throughout the bulk, defect bands are well suited to explain our bulk transport. Moreover, it explains the rather small band gaps observed by macroscopic experimental methods. Here, type 2 conductivity may have masked the actual bulk band gap, resulting in the measurement of an apparent band gap with reduced size. It is worth noting, that this is not in conflict with the reported band dispersion [16,30], as the observed spatially extended defect structures result in rather localized states in k-space.

As defect-related conductivity freezes out for lower temperatures, type 2 conductivity is not suitable to explain the conductivity of the freshly cleaved surface, as seen in figure 1(d). Here, the dependence of the resistance on the temperature changes drastically below 200 K, resulting in a plateau where the resistance becomes up to an order of magnitude lower compared to the bulk. This resemblance to metallic (surface-)states can be suitably explained with the type 3 conductivity, that is showing the spectroscopic signature for a band structure with a gap-closing via a linear 2D dispersion, i.e. resembling the vicinity of a Dirac-cone. For type 3 spectra, deviations from this V-shape are only found at the onset of tunneling into the valence and conduction bulk band states and beyond. Such a linear dI/dV is commonly not associated with defect states. This leaves either room for some form of exotic surface state of unknown physical origin as discussed in a recent ARPES work [30] or e.g. the theoretically discussed topologically protected surface conductivity in honeycomb iridates. Ref. 11 ruled out nontrivial topology since the corresponding Dirac cone was found at the $\bar{\Gamma}$-point instead of the $\bar{M}$-point in the k-space, contradicting theoretical calculations for 2D topological nontriviality of $Na_2IrO_3$ [30]. The possibility of 3D topologically insulating behavior as well as possible defect bands to explain the observed in-gap states were not considered [30].

## V. ROBUSTNESS OF SURFACE CONDUCTIVITY ACROSS STEP EDGES

To further investigate the physical origin of the type 3 spectral signature in the context of possible topological non-triviality, we have investigated the spatial dependence across specific defects like step edges. This is analogous to other studies that have investigated topological materials on a local scale [27]. Topologically protected conductivity can be categorized by dimensionality, resulting in 1D conducting channels at step edges or phase boundaries for 2D topological insulators and 2D conducting surfaces for 3D topological insulators [22-24]. The lateral resolution of STM/STS methods provides the opportunity to determine the dimensionality of conductive channels by investigating corresponding surface structures, e.g. step edges and free surfaces. An STM/STS-dataset covering a single step edge between two (1x1)-reconstructed surfaces is depicted in figure 4. Although the overall in-gap conductivity varies on the two terraces and across the step, all spectra resemble the V-shaped spectra in figure 3(c) qualitatively, i.e. are predominantly of type 3. Since no increased conductivity is found at the very step edge, we conclude that this V-shaped linear-dispersion in-gap conductivity (type 3) is mediated by the free surfaces and robust over step edges.

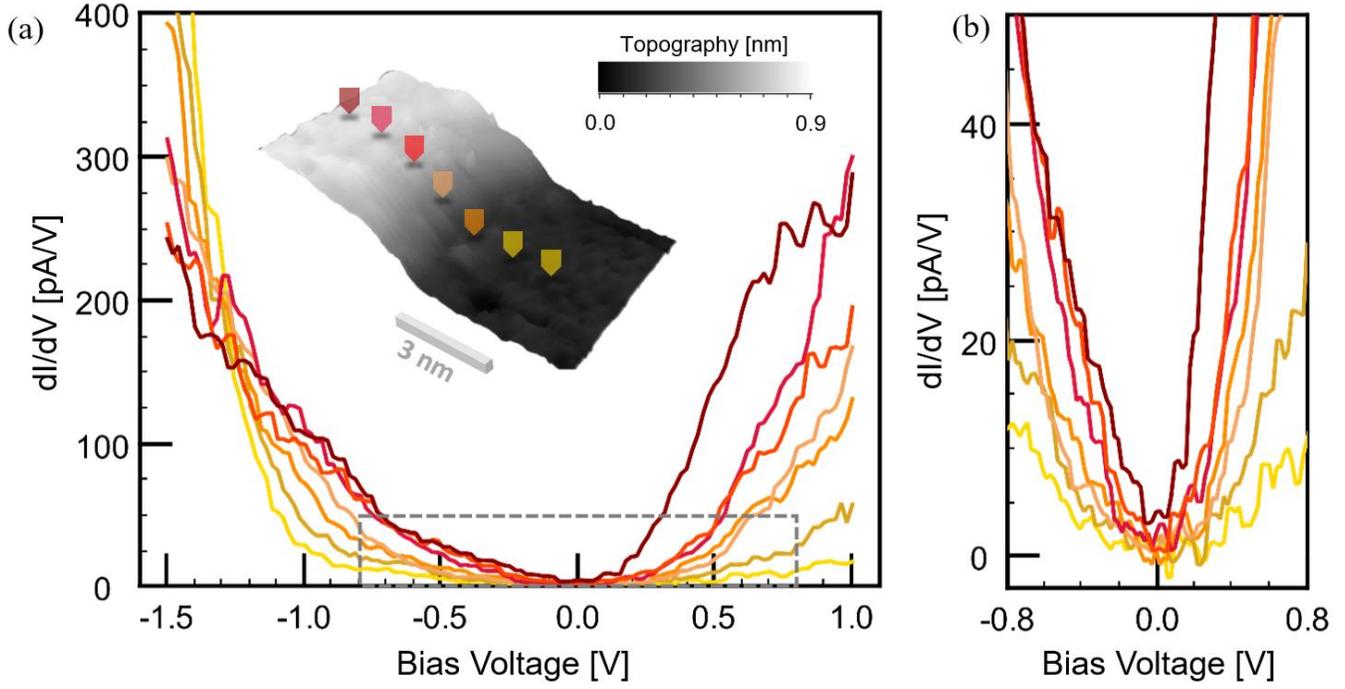

FIG. 4. Impact of step edges on type 3 spectra. (a) dI/dV-spectra recorded equidistantly across a single step on a (1x1)-reconstructed surface. The lateral distance between the spectra is 1 nm. The actual recording positions are marked in the inset and color-coded according to the dI/dV-curves. A smooth transition of the curves is observed when crossing the step edge, without any change in the qualitative shape of the spectra. However, the overall conductivity is altered. The deviation regarding the valence band onset at bias voltages lower than -0.8 V is due to the fact that the integral of the dI/dV-curves from the setpoint to the fermi-level has to be constant. (b) Zoom-in according to the grey dashed box in (a). The band gap closing resembles a V-shape on an energy scale of several hundred meV.

Combining the results from macroscopic temperature-dependent transport with the local information, we propose the following model. Bulk conductivity at higher temperatures is determined by thermally activated carriers from defects-related band-like states to the conduction band and at lower temperatures by hopping between these states, where both processes freeze out with decreasing temperature. This idea is supported by the fact, that the experimentally determined activation energy of ~200 meV fits to the spectroscopic energies found in STS spectra of type II. In contrast to former experimental studies, we find a clear signature of a low resistance surface conductivity with a sheet resistance between 0.5 k$\Omega$/□ and 2 k$\Omega$/□, comparable to a Si(111)7x7-surface (~1.25 k$\Omega$/□ at 300 K) [35] and Graphene on 6H-SiC (~0.6 k$\Omega$/□ at 200 K and ~0.8 k$\Omega$/□ at 300 K) [36]. This is an unexpectedly low value considering the high defect density at the $Na_2IrO_3$ surface (fig. 1(g)) and the relatively poor bulk conductance at low temperatures. Combined with the surprising robustness against dosing the surface with air, this supports the idea of a rather unconventional surface conductivity.

In addition, STS shows the spectroscopic signature of a linear dispersion of a 2D system, which is robust across local defects like steps on the surface. As the theoretical modeling of honeycomb iridates opens up the possibility of nontrivial topology [9-11,20,21], it is tempting to follow the idea of $Na_2IrO_3$ being a 3D topological insulator. This implies that the V-shaped linear dispersion in gap conductivity is the fingerprint of possible protected states of 3D topologically non-trivial $Na_2IrO_3$. We have first indications that the main findings regarding the in-gap conductivity seem to be similar for the (1x1)- and the ($\sqrt{3}$x$\sqrt{3}$)R30°-surface (see Supplementary Material [Publisher-URL] for fig. S6.), supporting the prospect of $Na_2IrO_3$ being 3D topologically non-trivial. Consequently, the premise for the interpretation of previously observed surface metallicity [30] and the resulting rejection of topological non-triviality for $Na_2IrO_3$ needs to be reconsidered.

At this point, we would like to speculate that the conductive, and therefore potentially topologically protected, layer may be not located directly at the very surface but might have a small distance to the surface. This idea is motivated by two experimental observations: Firstly, the conductivity is rather robust against air. Even after dosing the surface with air for several hours the surface conductivity is still partially present. Secondly, to address the type 3 conductive channel, we need a specific combination of the electronic structure of the surface as well as the tip. Moreover, it might explain the difficulty to overserve the related states by means of ARPES.

## VI. CONCLUSION

We report the first experimental evidence of a TMO to show a hallmark of topological insulator behavior with a huge bulk band gap of 0.8 eV that is comparable to common semiconductors. The combination of electronic properties of the STM-tip, as well as the local surface structure has a severe impact on the dI/dV-spectra, bringing the addressability of different states into the focus. Three basic accessible conducting channels are identified: regular bulk conductance (type 1) for voltages greater than the bulk band gap, conductance via defect bands (type 2) and presumably topologically protected surface conductance (type 3). Future investigations of the nature of the addressability of electronic states as well as accompanying theoretical work are key to clarify the origin and nature of topological insulator behavior in $Na_2IrO_3$.

A topological insulator with a bulk band gap of 0.8 eV would be a leap towards the technological use of protected surface states, resulting in the prospect of 2D electronics with monoatomic thickness. Here, the class of honeycomb iridates represents a highly promising type of material, particularly deserving future research effort. In fact, our work shows that $Na_2IrO_3$ is a promising candidate for a large band gap transition metal oxide topological insulator, potentially bridging the fields of topology and strongly correlated systems.

Supplement

## CRYSTAL STRUCTURE OF NA2IRO3

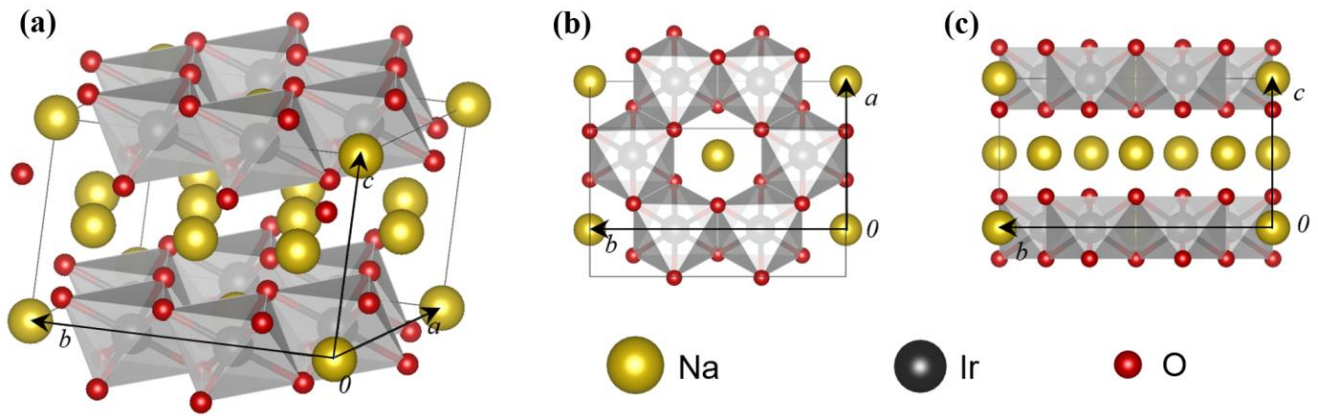

FIG. S1. Crystal structure of $Na_2IrO_3$. (a) Unit cell of $Na_2IrO_3$, consisting of alternatingly stacked honeycomb $NaIr_2O_6$ and $Na_3$ layers [14,31]. (a) Top view of the $NaIr_2O_6$ layer, showing the edge-sharing $IrO_6$ octahedra forming a honeycomb lattice. The center of every honeycomb is occupied by a single Na atom. (c) Side view of the $Na_2IrO_3$ layer structure. The respective topmost (surface) layer lacks 2/3 of sodium after cleavage [15].

# ATOMICALLY RESOLFED SURFACE STRUCTURE OF THE (1x1)-RECONSTRUCTED NA2IRO3-SURFACE

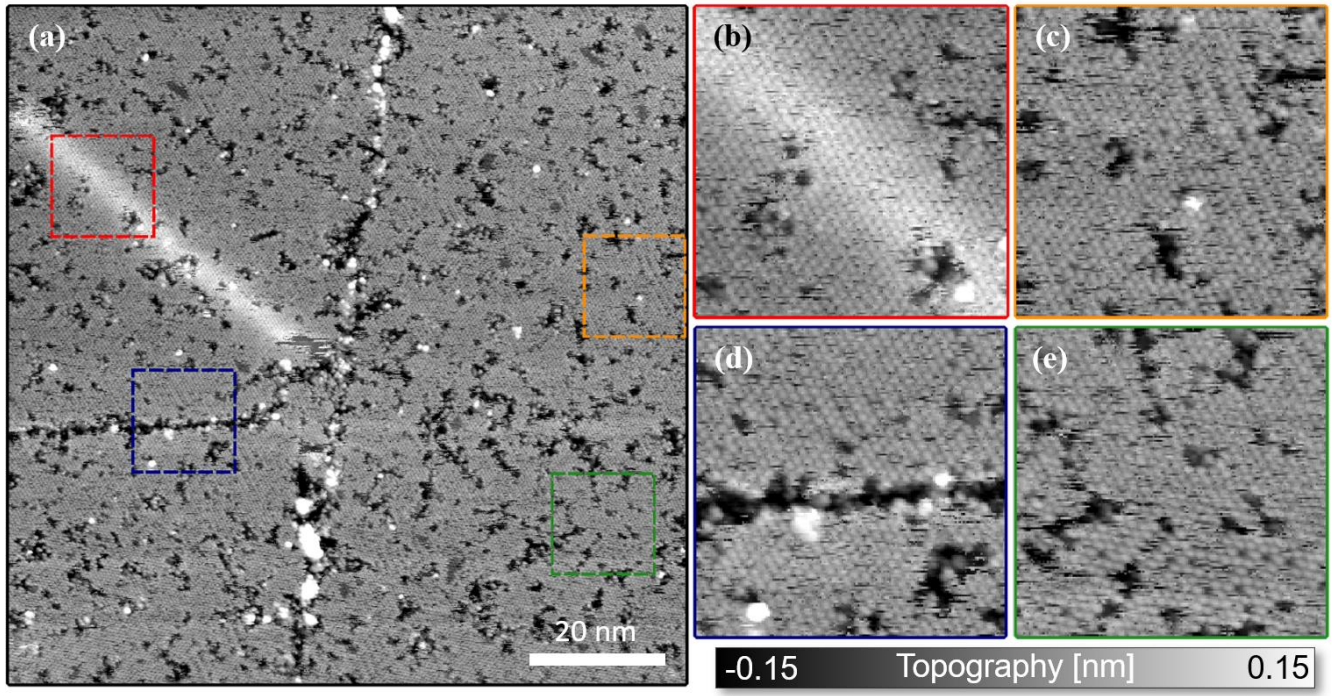

FIG. S2. High defect density observed on $Na_2IrO_3$ surfaces. (a) 100 nm x 100 nm constant current topographic map showing various defect structures on the (1x1)-reconstructed $Na_2IrO_3$ surface, recorded using a setpoint of -1.5 V and 100 pA. The dashed red boxes mark the positions of the zoom-ins depicted in (b)-(e). (b) Ridge-like defect with a topographic height of approximately 100 pm. (c) "Stripy" pattern superimposed to the regular (1x1)-reconstructed surface, bordered by various defects. The pattern does not extend over any defects. (d) Crack in the topmost crystal layer (here: $Na_3$ layer). Such cracks extend over several hundred nanometer and locally exhibit small sodium aggregations. (e) Different surface defects, ranging from vacancies due to missing single sodium atoms to large defect aggregations.

# IMPACT OF STM-TIP MODIFICATIONS ON THE TOPOGRAPHY RESOLUTION AND SPECTROCOPIC DATA

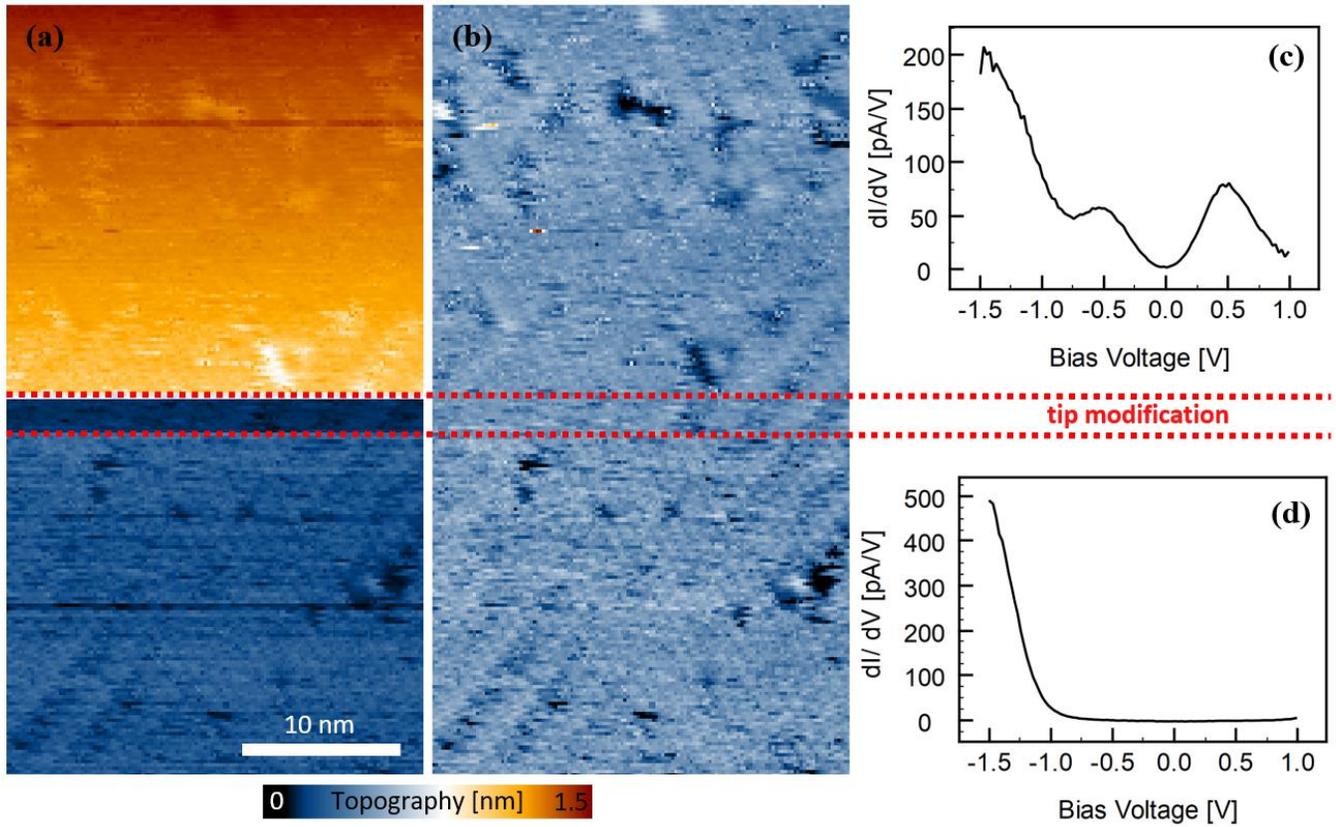

FIG. S3. Maintaining atomic resolution despite tip modifications. (a) Constant current topography of the (1x1)-reconstructed surface recorded using a setpoint of -1.5 V and 100 pA. Two consecutive tip modifications occur in the center of the topography, indicated by abrupt jumps of the topographic height. (b) Same topography as in a with adjusted height to smoothen the topographic jumps caused by the tip modifications. (c) Average dI/dV spectrum before the tip-modification. (d) Average dI/dV spectrum after the tip-modification.

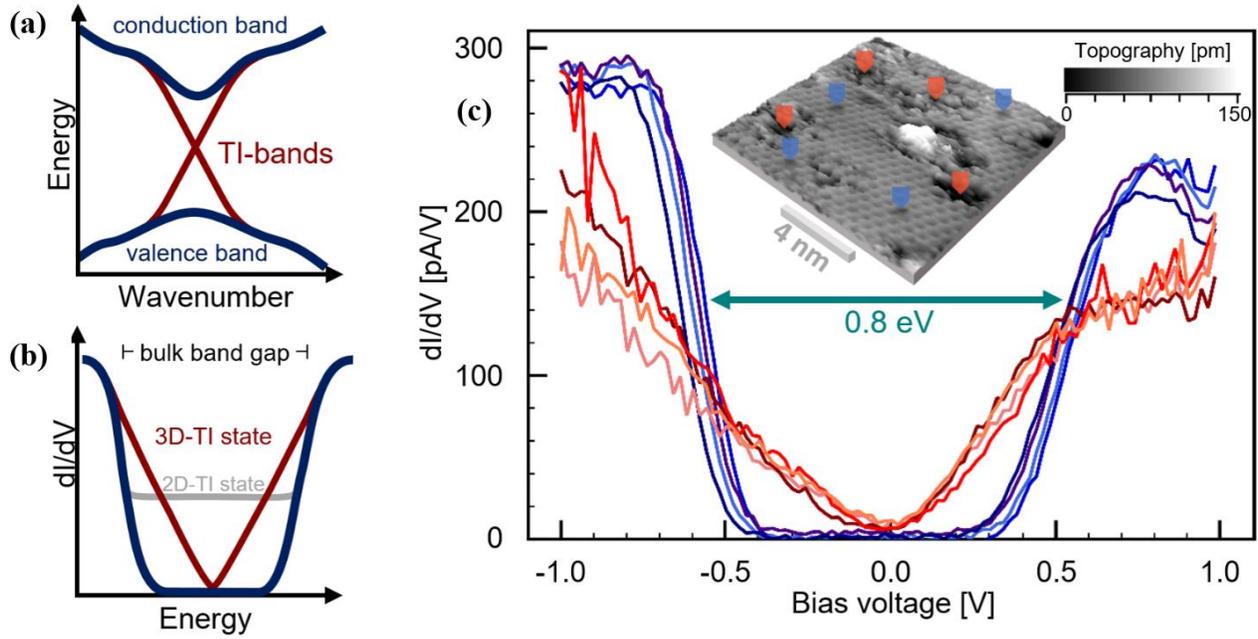

FIG. S4. Dirac-like spectral density evidencing 3D-TI behavior in the strongly correlated $Na_2IrO_3$. (a) Scheme of the linear dispersion (Dirac-cone) band gap closing due to the bulk-boundary correspondence [23] for interfaces between materials with different topological index. (b) Expected dI/dV-curves for a 3D- and a 2D-topological insulator within the bulk band gap region. As 2D-TI's exhibit protected 1D conducting channels with a linear band structure, the corresponding density of states must be constant. 3D-TI's on the other hand have 2D conducting surfaces with a Dirac-like band structure, resulting in a V-shaped density of states. Since dI/dV curves are proportional to the local density of states, the same expectations regarding the shape of the curves hold for the measured dI/dV spectra. (c) dI/dV spectra at different positions on the 1x1 reconstructed $Na_2IrO_3$ surface. Blue spectra with a bandgap of 0.8 eV were taken at clean, intact surfaces. The red spectra, exhibiting a band gap closing in a V-shaped fashion, were recorded at vacancies. Tunneling into the valence and conduction bulk band results in a kink at -0.45 V and 0.35 V respectively. The exact recording positions of the spectra are marked in the inset, taken with at a setpoint of -1 V and 100 pA.

# VARIATION OF CONDUCTIVITY CHANNELS IN DEPENDENCE OF THE TIP AND SURFACE STRUCTURE

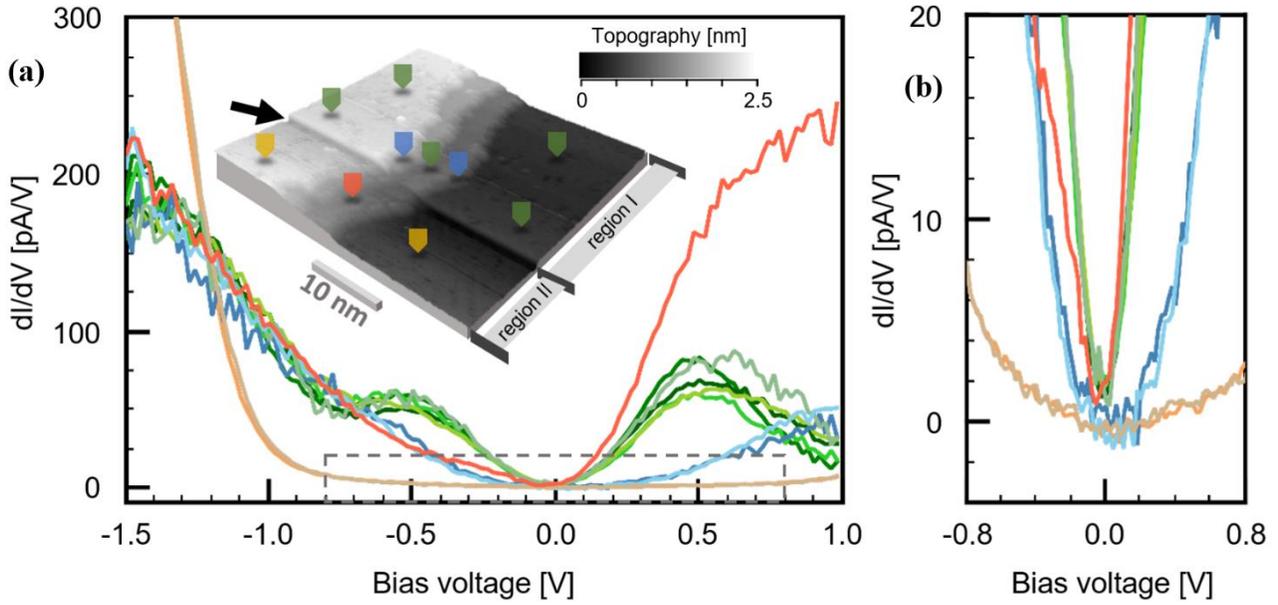

FIG. S5. Access to different conductivity channels after tip-modification. (a) dI/dV-spectra at terraces and step edges before and after the disposal of tip-attached sodium, according to the positions marked in the inset. Spectroscopy measurements were performed with an initial setpoint of -1.5 V bias voltage and 100 pA tunneling current at every second pixel in the topography. The jump in topographic height caused by the tip modification is marked by a black arrow. We refer to the area probed with the sodium modified tip as region I. Region II covers the probed area after the tip modification. (b) Zoom-in of the spectroscopic data according to the dashed grey box in (a).

# FIRST INIDICATION OF A CLOSED BAND GAP ON THE (√3 x √3)R30°-RECONSTRUCTED SURFACE OF NA2IRO3

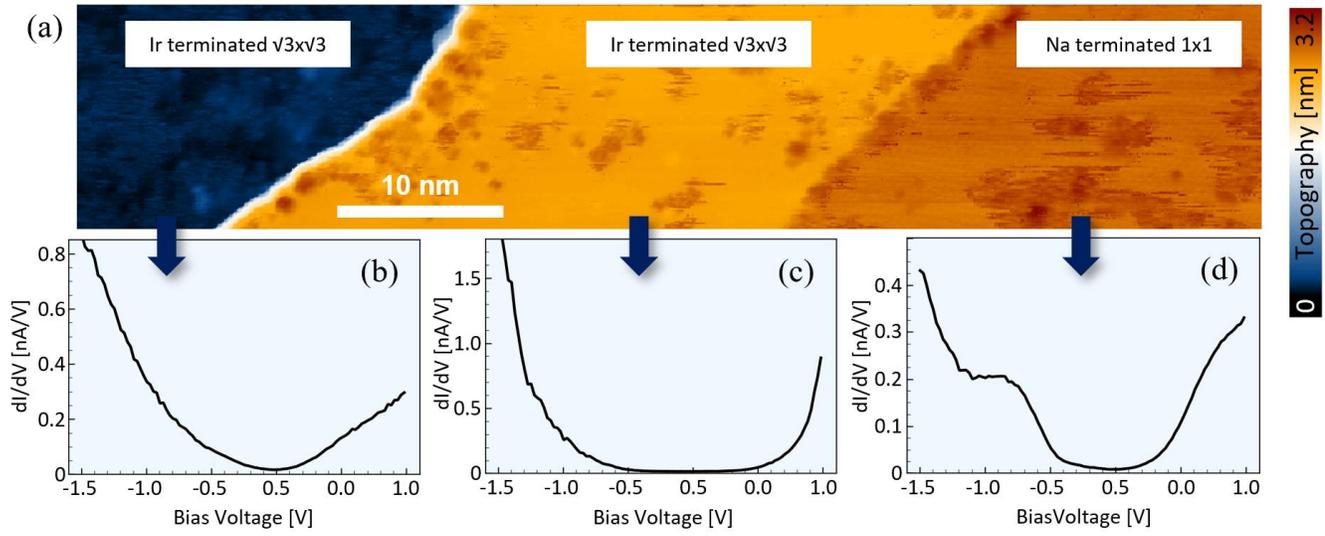

FIG. S6. Simultaneous observation of an open band gap as well as a V-shaped band gap closing for the (√3x√3)R30°-reconstructed surface. (a) Constant current topographic map covering one (1x1)-terrace and two (√3x√3)R30°-terraces, recorded using a setpoint of -1.5 V and 100 pA. (b)-(d) dI/dV-curves taken at the terraces corresponding to the marking with blue arrows.